\documentclass[showpacs,preprintnumbers,amsmath,amssymb]{revtex4}

\newcommand{\be}{\begin{equation}}
\newcommand{\en}{\end{equation}}
\newcommand{\bea}{\begin{eqnarray}}
\newcommand{\ena}{\end{eqnarray}}

\begin{document}
\title{Randall-Sundrum model with $\lambda<0$ and bulk brane viscosity}
\author{ Samuel Lepe$^{1}$, Francisco Pe{\~n}a,$^{2}$  and
Joel Saavedra$^{1}$}
\address{$^{1}${Instituto de F\'{\i}sica, Pontificia
Universidad Cat\'olica de Valpara\'{\i}so, Casilla 4950,
Valpara\'{\i}so, Chile.}\\
 $^{2}${Departamento de Ciencias F\'{\i}sicas, Facultad de
Ingenier\'{\i}a, Ciencias y Administraci\'on,\\ Universidad de la
Frontera, Avda. Francisco Salazar 01145, Casilla 54-D, Temuco,
Chile.}}
\date{\today}

\begin{abstract}

We study  the effect of the inclusion of bulk
 brane viscosity on brane world (BW) cosmology in the framework of the
 Eckart's theory, we focus in the Randall-Sundrum model
 with negative tension on the brane.
\end{abstract}


\maketitle

\section{\label{sec:level1} Introduction}

Extra dimensions theories have a long history, from the original
work of Kaluza-Klein \cite{kaluza} to modern ideas of string
theory \cite{Font:2005td, Chan:2000ms}. In particular the
Randall-Sundrum scenario has got great attention in the last
decade \cite{Randall:1999ee, Randall:1999vf}. From the
cosmological point of view, brane world offers a novel approach to
our understanding of the evolution of the universe. The most
spectacular consequence of this scenario is the modification of
the Friedmann equation. In these models, for instance in five
dimensions, matter is confined to a four dimensional brane, while
gravity can be propagated in the bulk,and can feel the extra
dimension. From the perspective of string theory ~\cite{witten},
brane world cosmology has been a big challenge for modern
cosmology . For a review on BW cosmology see Ref.~\cite{lecturer}.
For example, consequences of a chaotic inflationary universe
scenario in a BW model was described~\cite{maartens}, where it was
found that the slow-roll approximation is enhanced by the
modification of the Friedman equation. Many works in these
theories has been made considering a perfect fluid, and inclusion
of imperfect fluid has been less considered in comparison with the
former case, see for example Refs \cite{Brevik:2005gx,
Chakraborty:2005sn, Brevik:2003gh, Harko:2002hv, Chen:2001sma}. In
this work we study the effect of bulk viscous brane in
Randall-Sundrum model with negative tension ($\lambda<0$) in a
modified Friedmann-Lemaitre-Robertson-Walker (FLRW) model over the
brane. Effects of the imperfect fluids in the brane world
cosmology produce many differences with respect to standard
cosmology scenarios.

The plan of the paper is as follows: In Sec. II we specify the
effective four dimensional cosmological equations from a five
dimensional Anti de Sitter brane world  model. We write down the
cosmological equations for an imperfect fluid. Also, we discussed
the condition and consequences of thermodynamics equilibrium.
Finally, we conclude in Sec. III.

%%%%%%%%%%%%%%%%%%%%%%%%%%%%%%%%%%%%%%%%%%%%%%%%%
%                                               %
%           viscocidAD y Eckart                 %
%                                               %
%%%%%%%%%%%%%%%%%%%%%%%%%%%%%%%%%%%%%%%%%%%%%%%%%

\section{\label{sec:level2}Bulk viscosity and the Eckart's theory in Randall-Sundrum
Cosmological Scenario}

One of the most spectacular consequences of the cosmological brane
world scenario is the modification of the Friedmann equation, that
has allowed to study some cosmological puzzles from a new
perspective. The particular case of a five dimensional model and
matter confined to a four dimensional Brane, has been acquiring
great attention in the last time, for a review see Ref. \cite{rm}
and references therein. We are going to consider an homogeneous
and isotropic 4-brane described by the FLRW metric, in the case of
flat universe where the gravitational sector of the field
equations is described by a modified Friedmann equation given by
\begin{equation}
3H^{2} =\rho \left( 1\pm \frac{\rho }{2\lambda }\right) ,
\label{eq.1}
\end{equation}
where was considered that $8\pi G=1$, the positive and negative
signs are related to positive and negative brane tension. The
matter content sector satisfy
\begin{equation}
\dot{\rho }+3H\left( \rho +p \right) =0.  \label{eq.2}
\end{equation}
These equations show that just the gravitational field can escape
from the brane and propagate through the bulk, and that matter is
confined on the brane. In order to study the effect of a viscous
fluid on the brane, we include a viscous pressure on the
continuity equation
\begin{equation}
\dot{\rho }+3H\left( \rho +p+\Pi \right) =0,  \label{eq.2pi}
\end{equation}
where $\Pi $ represents viscous pressure over the brane. To show
some virtues of this model, we first review the case with $\Pi
=0$, $p=\omega \rho $ and $\omega<-1$ that was studied by
Srivastava in Ref.\cite{Srivastava:2007fc}. From Eqs.(\ref{eq.1})
and (\ref{eq.2}), it is possible to obtain an exact solution for
the matter contents
\begin{equation*}
\rho \left( t\right) =\left\{ \mp \frac{1}{2\lambda }+\left[ \sqrt{\frac{1}{%
\rho _{0}}\pm \frac{1}{2\lambda }}+\sqrt{\frac{3}{4}}\left(
1+\omega \right) \left( t-t_{0}\right) \right] ^{2}\right\} ^{-1}.
%\label{eq.3}
\end{equation*}
In Ref. \cite{Srivastava:2007fc} the case of brane world model
with $\lambda > 0$ was discussed, that describes an accelerated
phantom universe that ends in a big-rip singularity in the time
\begin{equation*}
t_{s}=t_{0}+\sqrt{\frac{4}{3}}\frac{1}{1+\omega }\left[ \sqrt{\frac{1}{%
2\lambda }}-\sqrt{\frac{1}{\rho _{0}}+\frac{1}{2\lambda }}\right]
.
\end{equation*}
For a brane world model with $\lambda < 0$, a phantom universe
that begins with  an accelerated phase and then evolves towards a
decelerated phase was found.

For our model, when $\Pi \neq 0$, it is useful to define the
adimensional variable $x=\rho /2\lambda $ that allow to rewrite
Eq. (\ref{eq.1}) in the standard form

\begin{equation}
3H^{2} =\rho _{eff},  \label{eq.3}
\end{equation}
where the effective density is given by \be \rho _{eff} =2\lambda
x\left( 1\pm x\right). \label{effectiverho} \en  Taking the
derivative of Eq.(\ref{eq.3}) respect to cosmological time and
using Eq.(\ref{eq.2pi}), we obtain the dynamical equation \be
\dot{H} =-\frac{1}{2}\left[ \rho _{eff}+p_{eff}+\Pi _{eff}\right],
\label{hpunto} \en where the effective pressures are given by
\begin{eqnarray}
p_{eff} &=&2\lambda \left[ \omega x\left( 1\pm 2x\right) \pm
x^{2}\right] ,
\label{eq.6} \\
\Pi _{eff} &=&\Pi \left( 1\pm 2x\right),  \label{eq.7}
\end{eqnarray}
and the effective state equation is given by
\begin{equation}
\omega _{eff}=\frac{p_{eff}+\Pi _{eff}}{\rho _{eff}},  \label{eq.9}
\end{equation}
where
\begin{equation}
\omega _{eff}\left( x\right) =\frac{1}{1\pm x}\left[ \left( 1\pm 2x\right)
\left( \omega +\frac{\Pi }{2\lambda x}\right) \pm x\right] .  \label{eq.10}
\end{equation}
This equation has the Friedmann limit when $x <<1$ and standard
cosmology is recovered. It is straightforward to see in this case
that the effective
barotropic index is given by $\omega _{eff}\, \rightarrow\, \omega +\frac{\Pi }{%
2\lambda x}$ ,i.e., one recovers the standard four dimensional
general relativity with a bulk viscosity.  On the other hand, the
strong limit $x>>1$ is given by $\omega _{eff}\, \rightarrow
\,2\left( \omega + \frac{\Pi }{2\lambda x}\right) + 1$, here the
quadratic term in the energy density dominates over other terms
giving rise to a new kind of behavior for the Friedmann equation.
Acceleration or deceleration phases without viscosity has been
discussed in the literature \cite{Srivastava:2007fc,Zhang:2007yu},
and for the particular case of $\lambda<0$ it is known as the
bouncing brane \cite{Shtanov:2002mb, Hovdebo:2003ug}.

Assuming that our cosmological fluid is viscous and to
characterize this novel behavior on the evolution in the universe,
we are going to study our model in the context of non-causal
thermodynamics that is known as Eckart Theory \cite{eckart}.
Although this model present some causality warning, it is the
simplest alternative and has been widely considered in cosmology,
see for example
Refs.\cite{Colistete:2007xi,Brevik:2005bj,Zimdahl:1996ka} and
references therein. Because we assumed spherical symmetry, the
shear viscosity does not play any role, and only the bulk
viscosity has to be considered,
\begin{equation}
\Pi \left( \rho \right) =-3\xi \left( \rho \right) H.  \label{eq.13}
\end{equation}
From a thermodynamical point of view, in conventional physics
$\xi$ needs to be positive; this is a consequence of the positive
entropy change in irreversible processes, For $\xi \left( \rho
\right) $ we consider
\begin{equation}
\xi \left( \rho \right) =\xi _{0}\rho ^{s},  \label{eq.14}
\end{equation}
and the effective state equation (\ref{eq.9}) reads
\begin{equation} \omega _{eff}\left( x,s\right) =\frac{1}{1\pm
x}\left[ \left( 1\pm 2x\right)
\left( \omega -\sqrt{3}\xi _{0}\left( 2\lambda \right) ^{s-1/2}x^{s-1}\sqrt{%
x\left( 1\pm x\right) }\right) \pm x\right] .  \label{eq.16}
\end{equation}
For $\lambda<0$, it is well known that the universe exhibit
bouncing and turnaround phases. These phases arise from the
modifications to the Friedmann equation. A simple analysis of Eqs.
(\ref{effectiverho})-(\ref{eq.7}) and Eq. (\ref{eq.16}) with a
negative sign, allows to characterize the behavior of the brane
universe filled with a viscous fluid. For example, if we take
$s=1/2$ it is straightforward to find a novel behavior of
$\Pi_{eff}(x)$ from Eq.(\ref{eq.16}), note that in the case $s
\neq 1/2$ the same behavior is obtained. In the interval
$0<x<1/2$, $\Pi_{eff}(x)<0$ admits a genuine interpretation in
term of a bulk viscosity and therefore can be considerate as a
source for entropy generation. On the other hand, when $1/2<x<1$
it is a quite problematic to give a conventional physics
interpretation of $\Pi_{eff}(x)>0$, this point will be discussed
elsewhere. Moreover, the effective description shows the crossing
of phantom divide $\omega_{eff}<-1$, this region correspond to a
new phantom bounce, due to viscosity effects. In order to check
the thermodynamic equilibrium, we study the if the limit
$\Pi_{eff}/p_eff<<1$ is satisfied, here
\begin{equation}
\left|\frac{\Pi_{eff}}{p_{eff}}\right|=\sqrt{3}\xi
_{0}\left|\frac{\sqrt{1-x}(1-2x)}{\omega(1-2x)-x}\right|,\label{te3}
\end{equation}
in the interval $0<x<1/2$ where $\Pi_{eff}$ is a genuine
viscosity. We note that Eq. (\ref{te3}) satisfies the limit
$\Pi_{eff}/p_eff<<1$ if $-1<\omega<-1/3$, therefore Eq.
(\ref{te3}) satisfies the condition for thermodynamic equilibrium.
When $\omega$ is outside this interval it is not possible to
satisfy this condition.

For a model with $\lambda>0$, it is possible to extend the
previous analysis and to obtain an exact solution. First, we
rewrite the continuity equation (\ref{eq.2}) in terms of the
adimensional variable $x$
\begin{equation}
\dot{x}+\left( 1+\omega \right) x\sqrt{6\lambda x\left( 1+
x\right) }=3\xi _{0}\left( 2\lambda \right) ^{s}x^{s+1}\left( 1+
x\right) .  \label{eq.15}
\end{equation}
At early time in the evolution of the universe, where the
viscosity ($\Pi $) could be relevant, we consider the quadratic
correction as the dominant term in the Friedmann Eq. (\ref{eq.1})
\cite{Zhang:2007yu}. Thus, Eq. (\ref{eq.15}) can be integrated as
follow
\begin{equation}
t\left( x\right) =\frac{1}{B}\left( \frac{1}{x}+\int dx
\frac{x^{s-2}}{x^{s}-B/A}\right)   \label{eq.19}
\end{equation}
where the constants $A$ and $B$ are given by
\begin{equation}
A=3\xi _{0}\left( 2\lambda \right) ^{s}\text{ \ \ }\,and\,\text{ \ \ }B=\sqrt{%
6\lambda }\left( 1+\omega \right) .  \label{eq.20}
\end{equation}
%\LerchPhi{z}{s}{a}
%the Lerch transcendent Phi(z, s, a).
An explicit solution of Eq. (\ref{eq.19}) is given by
\begin{equation}
t\left( H\right) =-\frac{1}{3\left( 1+\omega \right) }H^{-1}\frac{1}{s}%
\,\,\Phi( \frac{3^{\left( s+1\right) /2}\xi _{0}\left( 2\lambda
\right) ^{\left( s-1\right) /2}}{\left( 1+\omega \right) }H^{s},1,-\frac{1}{s%
}),   \label{eq.21}
\end{equation}
where the Lerch transcendent $\Phi(z, s, a)$ function is defined
by,
\begin{equation}
\Phi(z,s,a)=\sum\limits_{n=0}^{\infty}\frac{z^n}{(n+a)^{s}}.
\label{eqler}
\end{equation}

 $\Pi_{eff}<0$ in the interval allowed for $x$ and it is
 straightforward to verify that the condition for the thermodynamic equilibrium is not
satisfied. A complete qualitative description for model with
$\lambda>0$ in the context of causal thermodynamics can be found
in Ref. \cite{Chen:2001sma}. We want to finish this section with a
comparison of our formalism with the structure that was obtained
in the context of loop quantum cosmology with $k=0$ \cite{lqc} the
modified Friedmann equation can be written as,
 \be 3H^{2}
=\rho \left( 1-\frac{\rho }{\rho _{c}}\right) ,  \label{eq.25} \en
where $\rho_c$ is the critical energy density set by quantum
gravity. This scheme including viscosity admits the same analysis
and shows some results quite similar to brane world cosmology in
the presence of viscosity for $\lambda<0$.

\section{\label{sec:level3}Discussion}
In this work we have studied the effect of bulk viscous brane in
Randall-Sundrum model. For non-causal thermodynamics, in the
context the Randall-Sundrum model with $\lambda<0$ , we have
considered the viscous pressure $\Pi \left( \rho \right) =-3\xi
_{0}\rho ^{s} \left( \rho \right) H$, and we have shown that in
the interval $0<x<1/2$, $\Pi_{eff}(x)<0$ admits a genuine
interpretation in terms of bulk viscosity. Because
$\Pi_{eff}(x)<0$ we can interpret $\Pi_{eff}(x)$ as a source for
entropy generation. Besides, the effective description shows the
possibility to produced crossing of phantom divide
$\omega_{eff}<-1$, this region corresponds to a new phantom bounce
due to viscosity effects. Through of the study of the ratio
$\Pi_{eff}/p_eff$, in order to check the thermodynamics
equilibrium, we have shown that it is possible to satisfy
$\Pi_{eff}/p_eff<<1$ . On the other hand, when $1/2<x<1$, it is
quite delicate to give a interpretation for $\Pi_{eff}(x)>0$ in
terms of conventional physics. We hope to clarify this point in
the near future. For a Randall-Sundrum model with $\lambda>0$, we
had extended the previous analysis and found an implicit exact
solution for the Hubble parameter. We also noted that
$\Pi_{eff}<0$ in all interval allowed for the variable $x$ and the
condition of the thermodynamic equilibrium is not satisfied.

\begin{acknowledgments}
This work was supported by COMISION NACIONAL DE CIENCIAS Y
TECNOLOGIA through FONDECYT \ Grants 1040229 (SL) and 11060515
(JS). This work was also partially supported by PUCV Grants No.
123.792/2007 (SL), No. 123.789/2006 (JS), and from DIUFRO No
120618 of Direcci\'on de Investigaci\'on y Desarrollo Universidad
de la Frontera (FP) . The authors SL and JS want to thank to
Departamento de F\'{\i}sica de la Universidad de La Frontera for
its kind hospitality.
\end{acknowledgments}

\end{document}